\documentclass{ragtime} % NO optional arguments are required
\title[Tidal deformability of ultracompact Schwarzschild stars]%
      {Tidal deformability of ultracompact Schwarzschild stars and their approach to the black hole limit}

\author[C. Posada % Run. head authors: separate names with commas,
            ]%    Now let's start the paper title authors:
       {Camilo Posada\at{1} % Makes referencing superscript `1'
        \\% Termination of authors' block; if
        \ins{1}Institute of Physics in Opava,
        Silesian University in Opava,\splitins[1]% This is how to break an
        Bezru\v{c}ovo n\'am.~13, CZ-746\,01 Opava,
        Czech Republic\\% Termination of the first affiliation.
        \ins{a}\Email{camilo.posada@physics.slu.cz}} % This is how to present E-mail.

\coentry{C. Posada}%
       {}

\providecommand{\sch}{\mathrm{S}} % Schwarzschild radius
\providecommand{\rme}{\mathrm{e}} % Schwarzschild radius

\begin{document}

% Citation of references in abstract should generally be avoided to
% ensure self-consistency of the abstract.  If you do insist on citation(s)
% within the abstract, you should use the \bibentry command, which forces
% the _complete_bibliographic_entry_ to appear in the abstract.
% With the `nonatbib' optional class argument this feature is not available.
\begin{abstract}
A well-known result in general relativity is that the tidal Love numbers of black holes vanish. In contrast, different configurations to a black hole may have non-vanishing Love numbers. For instance, it has been conjectured recently that the Love number of generic exotic compact objects (ECOs) shows a logarithmic behavior. Here we analyze the ultracompact Schwarzschild star which allows the compactness to cross and go beyond the Buchdahl limit. This Schwarzschild star has been shown to be a good black hole mimicker, moreover, it has been found that the Love number of these objects approaches zero as their compactness approaches the black hole limit. Here we complement those results, by showing that the Love number for these configurations follows an exponentially decaying behavior rather than the logarithmic behavior proposed for generic ECOs. 
\end{abstract}

% The key words are to be separated by the N-dash surrounded by spaces,
% the left one non-breakable.  The name concatenation like Kerr--de~Sitter
% should also be typed with N-dash but with no spaces around (compare,
% e.g., Levi-Civita, which is a single person):
\begin{keywords}
Tidal deformability~-- Interior solutions~-- black hole mimicker~-- gravastar
%test particle~-- cosmological constant~-- spin dynamics~-- equilibrium
\end{keywords}

% It is good to provide as many as \label's possible, but never start the
% key with a numeral.  This makes problems with pdflatex processing.
\section{Introduction}\label{intro}%%%%%%%%%%%%%%%%%%%%%%%%%%%%%%%%%%%%%%%%

Black holes (BHs) are one of the most intriguing predictions of Einstein's classical general relativity (GR). Beyond their elegant mathematical structure \citep{Chandrasekhar:1985kt}, they are also the main candidates to explain most of the astrophysical observations \citep{Celotti:1999tg}. Nevertheless, besides their curvature singularity in the interior geometry, which is believed will be ‘removed' by a consistent quantum theory of gravity, BHs present certain paradoxes which remain puzzling. For instance, the interior geometry of mathematical BHs show some unphysical behavior, such as the closed timelike geodesics in the Kerr spacetime \citep{Hawking:1973uf}. On the other hand, the event horizon is at the root of the so-called ‘information paradox' which remains as one of the central problems in black hole physics \citep{Wald:1999vt}.

As a consequence of the BH paradoxes, a number of models of ‘regular BHs' or ‘BH mimickers' (also known as exotic compact objects (ECOs)) have been proposed in the literature (see e.g. \citep{Cardoso:2019rvt} and references therein). Regular BHs are constructed with different non-singular interiors, but in such a way that they reproduce the well-known geometries of BHs solutions in GR. For instance, the gravitational condensate star, or gravastar \citep{Mazur:2001fv, Mazur:2004fk}, is composed of a non-singular de Sitter interior with negative pressure $p=-\epsilon$, but with a positive vacuum energy $\epsilon>0$, which is matched to the exterior Schwarzschild solution with $p=\epsilon=0$. The ‘gluing' of both geometries was done by introducing an infinitesimal thin-shell of ultra-stiff matter.  

In connection with gravastars, \cite{Mazur:2015kia} revisited the well known Schwarzschild's interior solution with uniform density, or \emph{Schwarzschild star} (see e.g. \citep{Glendenning:1997wn}). The Schwarzschild star manifests a divergence in the central pressure when its compactness reaches the Buchdahl bound $M/R=4/9$. The importance of this limit relies on the fact that it is independent of the equation of state (EOS) of the configuration, as shown by \citep{Buchdahl:1959B}, under the assumption of isotropic pressure, positive energy density and monotonically decreasing with the distance $r$. Thus the Schwarzschild star represents a toy model which saturates the Buchdahl bound and should be considered as the limiting case of an ultra-stiff EOS; incidentally, configurations with compactness higher than $4/9$ have been usually assumed as unphysical.

Nevertheless, some interesting features become apparent when one considers the Schwarzschild star beyond the Buchdahl limit, $R_\sch<R<(9/8)R_\sch$, where $R_\sch\equiv 2M$. First of all, the pole where the pressure is divergent moves out from the origin up to a surface of radius $R_{0}=3R\sqrt{1-(4/9)(R/M)}<R$, and a regular interior region with negative pressure emerges naturally in the regime $0<r<R_{0}$; meanwhile the pressure remains positive in the region $R_{0}<r<R$. In the limit when $R_{0}\to R_{\sch}^{-}$, from below, and $R\to R_{\sch}^{+}$, from above, the ultracompact Schwarzschild star becomes essentially the gravastar proposed by \cite{Mazur:2015kia}. It is important to remark that the Schwarzschild star \emph{evades} the Buchdahl limit by having an anisotropic stress at the surface $R_0$. In the limiting case when $R_0=R=R_{\sch}$, the interior static de Sitter is matched to the exterior Schwarzschild geometry through a boundary layer located at their respective horizons $R_{\sch}=H$, where $H$ is related to the de Sitter energy density by $\epsilon=3H^2/(8\pi)$. Furthermore, there is a discontinuity $[\kappa]$ in the surface gravities which produces a surface tension $\tau_s=1/(8\pi R_{\sch})$ and incidentally a $\delta$-function stress tensor which replaces the BH horizon.  

Some of the physical properties and observational signatures of the ultracompact Schwarzschild stars have been studied recently in the literature. For instance, a time-dependent model was proposed by \cite{Beltracchi:2019tsu}. On the other hand, these configurations seem to be stable against radial oscillations \citep{Camilo:2018goy}. The analysis of axial modes was carried out in \citep{Konoplya:2019nzp}, where it was found that the Schwarzschild stars are stable against axial perturbations. Moreover, the Schwarzschild star can ‘mimic' very well the gravitational wave response of a BH at $l=2$ and higher multipoles, because it approaches the Schwarzschild BH spectrum as close as possible. This is due to the fact that the null surface $R_0$ provides the same boundary conditions for the quasi-normal modes as for the case of a BH. An extension to the anisotropic case using the minimal geometric deformation (MGD) was developed by \cite{Ovalle:2019lbs}.  

An early model for a slowly rotating ultracompact Schwarzschild star was proposed by \cite{Posada:2016qpz} using the Hartle-Thorne framework \citep{Hartle:1967he, Hartle:1968si}. However, the results reported there are marred by a wrong assumption, in the regime beyond the Buchdahl limit, after making a coordinate transformation. This proposal has been surpassed recently by \cite{beltracchi2021slowly} where they developed a model for a slowly rotating gravastar, to second order in the rotation. These authors found that the exterior metric to a slowly rotating gravastar is precisely that of the Kerr spacetime, therefore it's not possible to tell a gravastar from a Kerr BH by any observation, such as accretion disks processes or light ring images.   

An alternative for, potentially, distinguishing ECOs from BHS is through their tidal deformability. A compelling result in GR is that the tidal Love numbers of Schwarzschild BHs is zero \citep{Damour:2009vw,Binnington:2009bb,Hui:2020xxx,Chia:2020yla,Charalambous:2021mea,Poisson:2021yau}. On the other hand, it has been found that the Love numbers of general ECOs scale as $k\sim 1/\log\xi$, where $\xi$ is a parameter which measures how much the object deviates from the BH geometry \citep{Cardoso:2017cfl}. Recently \cite{Chirenti:2020bas} studied the tidal Love number of ultracompact Schwarzschild stars, below and beyond the Buchdahl limit. These authors found that the Love number of these configurations tends to zero as the compactness approaches the BH limit. Thus, they concluded that the vanishing of the Love number is not a unique property of BHS, instead, it's a consequence of the approach to the Schwarzschild limit. 

In this paper, we will review the main results of the tidal deformability of ultracompact Schwarzschild stars presented by \cite{Chirenti:2020bas}. As an addition to those results, we will show that the tidal Love number $k_2$ for ultracompact Schwarzschild stars do not follow the $1/\log\xi$ proposed by \citep{Cardoso:2017cfl}, instead, we found that $k_2$ decays exponentially as a function of the compactness. 
  
\section{Tidal deformability}
\label{sec:tidal}
\subsection{General formulation}
\label{sec:general}
The tidal Love number quantifies the deformations of the quadrupole moments of a star induced by external fields, which are connected through the relation \citep{Hinderer:2007mb,Damour:2009vw} 
\begin{equation}\label{quadrupole}
Q_{ij}=-\frac{2k_{2}R^5}{3}E_{ij}\equiv-\Lambda\,E_{ij}\,.
\end{equation}  
\noindent where $k_{2}$ is the Love number and $\Lambda$ is the tidal deformability. It is conventional to introduce the dimensionless tidal deformability
\begin{equation}\label{tidal}
\bar{\Lambda}=\Lambda/M^5=2k_{2}/(3C^5)\,,
\end{equation} 
here $C\equiv M/R$ denotes the compactness of the configuration. Following \cite{Damour:2009vw}, the unperturbed spacetime of a nonrotating star is described, generally, by the standard metric 
\begin{equation}
ds^2 = -e^{\nu(r)}dt^2 + e^{\lambda(r)}dr^2 + r^2d\Omega^2\,,
\end{equation}
and the even-parity metric perturbation $H=H_{0}=H_{2}$ is governed by the following equation
\begin{equation}\label{main}
\frac{d^2H}{dr^2}+C_{1}(r)\frac{dH}{dr}+C_{0}(r)H=0\,,
\end{equation}
where the coefficients $C_1$ and $C_0$ are given by
\begin{equation}\label{coef1}
C_{1}(r) = \frac{2}{r}+e^{\lambda}\left[\frac{2m}{r^2}+4\pi r(p-\epsilon)\right]\,, \\
\end{equation}
\begin{equation}\label{coef0}
C_{0}(r) = e^{\lambda}\left[-\frac{l(l+1)}{r^2}+4\pi(\epsilon+p)\frac{d\epsilon}{dp}+4\pi(5\epsilon+9p)\right] 
- \left(\frac{d\nu}{dr}\right)^2\,.
\end{equation}
In order to simplify the form of the perturbation equation \eqref{main}, it is conventional to introduce the logarithmic derivative $h(r) \equiv (r/H)dH/dr$. Substituting $h(r)$ into Eq.~(\ref{main}), we obtain a Riccati-type equation in the form \cite{Damour:2009vw}
\begin{equation}\label{ricatti}
r\frac{dh}{dr}+h(h-1)+rC_{1}h+r^2C_{0}=0\,,
\end{equation}
with the regular solution near the origin, $h(r)\simeq l$. Finally, the Love number $k_{2}$ can be determined using the following expression
\begin{equation}\label{k2}
\begin{aligned}
 k_{2}(C,h_{R}) = {} & \frac{8}{5}(1-2C)^2C^5[2C(h_R-1)-h_R+2]\times\{2C\big[4(h_R+1)C^4 \\
 & + (6h_R-4)C^3 +(26-22h_R)C^2 + 3(5h_R-8)C - 3h_R+6\big] \\
 &  +3(1-2C)^2\left[2C(h_R-1)-h_R+2\right]\log(1-2C)\}^{-1}.
%\label{k2}
\end{aligned}
\end{equation}
where $h_R$ is the value of $h$ at the surface $r = R$.

\subsection{Schwarzschild stars}
\label{sec:uniform}

In this section, we discuss the homogeneous configurations with constant density. Although uniform density stars are only an approximation for a realistic compact object, they are useful toy models which are described by Schwarzschild's interior solution \citep{Glendenning:1997wn}. Moreover, there are various reasons to considerer them in detail (see e.g. \cite{HTWW:1965}).

The tidal Love number for constant-density stars, or Schwarzschild stars, was studied by \citep{Damour:2009vw, Postnikov:2010yn,Chan:2014tva} and more recently by \citep{Chirenti:2020bas}. The details of the computation were discussed in these papers so we refer the reader to those works. 

The interior Schwarzschild solution describes a configuration of constant energy density $\epsilon$. For convenience, it can be written in terms of the auxiliary function $y$ \citep{Chandrasekhar:1974,Camilo:2018goy,Chirenti:2020bas} defined as
\begin{equation}
y^2=1-\left(\frac{r}{\alpha}\right)^2,\quad \textrm{with} \quad 
\alpha^2=\frac{3}{8\pi\epsilon} \equiv \frac{R^3}{R_{S}}\,,
\end{equation}
where $R$ is the radius of the star and $R_S \equiv 2M$ is the Schwarzschild radius. The function $y$ is defined in the range $[1, y_1]$ where $y_1 \equiv y(R)$ is the corresponding value at the surface. The interior metric functions are given by
\begin{equation}\label{sstar}
\rme^\nu=\frac{1}{4}(3y_{1}-y)^2\quad \textrm{and} \quad \rme^{\lambda}=\frac{1}{y^2}\,,\\
\end{equation}
which match continuously, at the surface $r=R$, with the exterior Schwarzschild solution. The pressure $p$ in the interior is found to be 
\begin{equation}
\label{p(y)}
p=\epsilon\left(\frac{y-y_{1}}{3y_{1}-y}\right)\,,
\end{equation}
which vanishes at the surface $r=R$. The central pressure diverges when the compactness reaches the Buchdahl bound $M/R=4/9$ \citep{Buchdahl:1959B}. 

It is convenient to introduce a new coordinate $x$ given by \citep{Chandrasekhar:1974}
\begin{equation}\label{xcoord}
x \equiv 1-y = 1-\sqrt{1-\left(\frac{r}{\alpha}\right)^2}\,,
\end{equation}
which is defined in the range $[0, x_1]$ where $x_1=1-y_1\equiv x(R)$, which depends on the compactness. The interior metric functions
\eqref{sstar} take the form

\begin{equation}\label{sstarx}
\rme^{\nu(x)}=\frac{1}{4}(k+x)^2, \quad  \textrm{and} \quad \rme^{\lambda(x)}=\frac{1}{(1-x)^2}\,,
\end{equation}
and the pressure (\ref{p(y)}) reads now
\begin{equation}\label{px}
p=\epsilon\left(\frac{1-x-y_1}{\kappa+x}\right),
%\label{eq:p(x)}
\end{equation}
where the constant $\kappa$ is defined as
\begin{equation}\label{k}
\kappa \equiv3y_{1}-1\,.
\end{equation}
Note that $\kappa>0$ when $\mathcal{C}<4/9$, and, $-1<\kappa\leq 0$ in the regime beyond Buchdahl. Note that Eq.~(\ref{px}), has a regular singular point at $x_0 \equiv -\kappa$. The careful analysis of this singularity, for the computation of the Love number, was done in \citep{Chirenti:2020bas} so we refer the reader to that paper for more details. 

The same singular point appears when one considers the extension of the ultracompact Schwarzschild star, to slow rotation, using the equations derived by Hartle \citep{Chandrasekhar:1974,Posada:2016qpz}. However, the results presented by \cite{Posada:2016qpz} are marred after assuming, wrongly, an absolute value of $\kappa$ when the compactness goes beyond the Buchdahl limit \footnote{The author acknowledges Emil Mottola and Philip Beltracchi for calling his attention to this point.}. 

\section{Tidal deformability of ultracompact Schwarzschild stars}\label{results}%%%%%%%%%%%%%%%%%%%%%%%%%%%%
In this section, we summarise the results of the tidal deformability of ultracompact Schwarschild stars reported by \cite{Chirenti:2020bas}. We used the transformation \eqref{xcoord} in the perturbation equation \eqref{ricatti}, which facilitates the computation when the compactness goes beyond the Buchdahl bound. Profiles of the solutions to the perturbation equation, for certain values of the compactness, are presented in Figs. 1 and 2 in \citep{Chirenti:2020bas}. Of particular interest is the value of $k_2$ at the Buchdahl limit which was found to be $\kappa_2^{\text{Buch}}=0.0017103$, which is in excellent agreement with the result reported by \cite{Damour:2009vw}. 

The value of $k_2$ is computed in the following way; the structure equations \eqref{sstarx}--\eqref{px} are substituted into the coefficients \eqref{coef1} and \eqref{coef0}, which consequently determine Eq. \eqref{ricatti}. Given the condition at the origin $h(r)= l$, one solves numerically Eq.~\eqref{ricatti} for $h(r)^{-}$, in the interior of the star, from the center (or, rather some cutoff value, $r_0=10^{-7}$), outwards up to the surface $r=R$. Some care must be taken here regarding the constant density condition. Note that Eq.~\eqref{coef0} contains a term in the form $(\epsilon+p)(d\epsilon/dp)$ which vanishes in the limit $\epsilon=\text{constant}$. Given the discontinuity of the density at the surface, this term contributes a $\delta-\text{function}$ which must be taken into account to obtain $h_{R}^{+}$ (see the discussion in \citep{Damour:2009vw,Postnikov:2010yn}). This correction is given by

\begin{equation}
h_{R}^{+}=h_{R}^{-}-\left(\frac{4\pi R^3 \epsilon}{M}\right)^{-}
\end{equation} 

\noindent which gives,

\begin{equation}
h_{R}^{+}=h_{R}^{-}-3
\end{equation}  

\begin{figure}[t]
\begin{center}
\includegraphics[width=\textwidth]{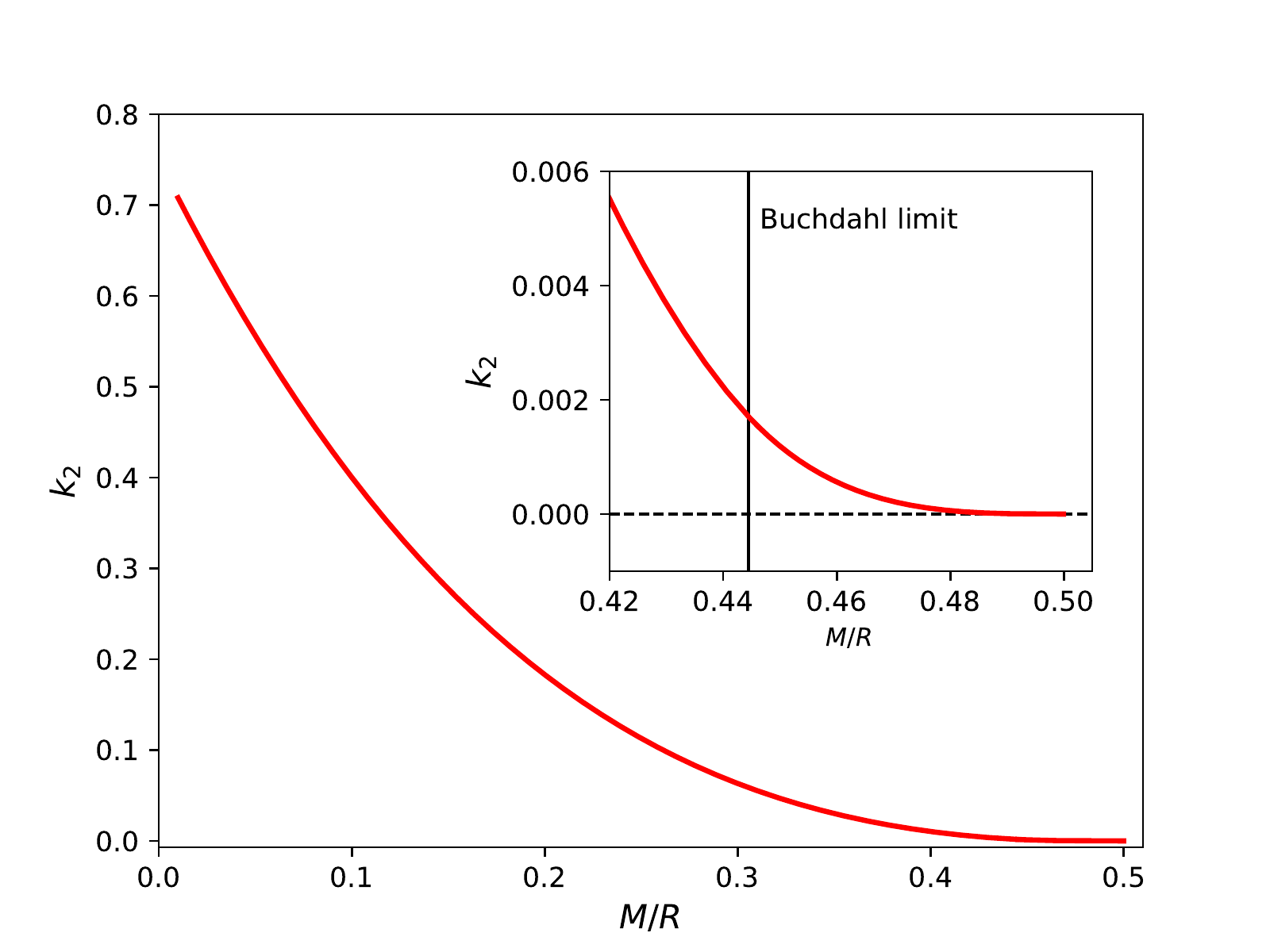}
\end{center}
\caption{Tidal Love number $k_2$, as a function of the compactness, for Schwarzschild stars below and beyond the Buchdahl limit. Note how $k_2$ approaches smoothly and continuously to zero, as the compactness approaches the Schwarzschild limit (figure adapted from \citep{Chirenti:2020bas}).} 
\label{Fig:1}
\end{figure}

In Fig.~\ref{Fig:1} we show the profile of the tidal Love number $k_2$, as a function of the compactness $M/R$, for Schwarzschild stars below and beyond the Buchdahl limit \citep{Chirenti:2020bas}. The inset shows a zoom of the region near the Buchdahl limit up to the Schwarzschild radius. The results of $k_2$ for $\mathcal{C}\leq4/9$ are in very good agreement with those reported by \cite{Damour:2009vw}. Note how $k_2$ approaches continuously to zero as the compactness approaches the Schwarzschild limit. Stars with compactness bigger than $4/9$ show a region of negative pressure in the interval $x\in[0,x_0]$ with $x_0=-\kappa$. In the limit when $\mathcal{C}\to 1/2$ from above, and $x_0\to x_1$ from below, the central region of negative pressure covers the whole interior of the star. Thus, in this limit, the ultracompact Schwarzschild star becomes essentially the gravastar \citep{Mazur:2015kia}. 

\begin{figure}[t]
\begin{center}
\includegraphics[width=\textwidth]{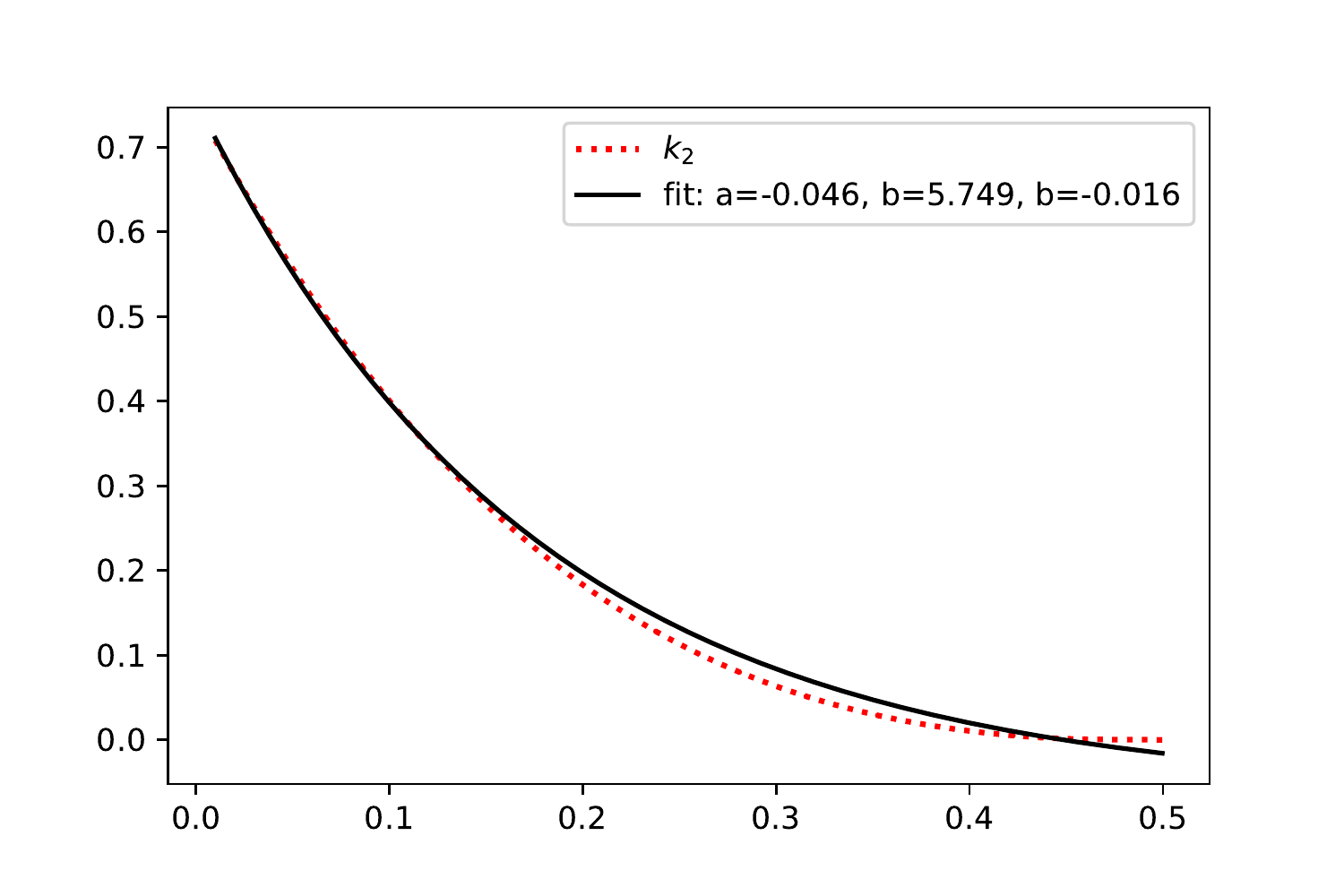}
\end{center}
\caption{Same results as in Fig. \ref{Fig:1}, but we also present the fit in a black solid curve. We consider an exponentially decaying model in the form: $a\left[1-e^{-b(\mathcal{C}-0.5)}\right]+d$. The corresponding coefficients are shown in the label.} 
\label{Fig:2}
\end{figure}

Note the striking ‘quenching', in three orders of magnitude, of the tidal Love number, for ultracompact Schwarzschild stars, from $0.75$, for $\mathcal{C}\to 0$, to $0.0017103$ at the Buchdahl bound, and the subsequent approach to zero as the compactness approaches $1/2$. This rapid quenching of $k_2$ clearly indicates that these configurations do not follow the logarithmic behavior suggested by \cite{Cardoso:2017cfl} as a ‘generic feature' of ECOs with an exterior geometry arbitrarily close to the BH geometry. To see this, in Fig.~\ref{Fig:2} we show the same results shown in Fig.~\ref{Fig:1} but we also include the fit which we model as an exponentially decaying function in the form
\begin{equation}\label{k}
k_2 = a\left[1-e^{-b(\mathcal{C}-0.5)}\right]+d,
\end{equation}
where the fitted coefficients are shown in the label of Fig.~\ref{Fig:2}. We found that the R-squared value for this fitting model is $R^2=0.999429$, which shows that the exponentially decaying model is a good one. Note that in reference \citep{Chirenti:2020bas} this fitting was overlooked so here we are complementing those results.

\begin{figure}[ht]
\begin{center}
\includegraphics[width=\textwidth]{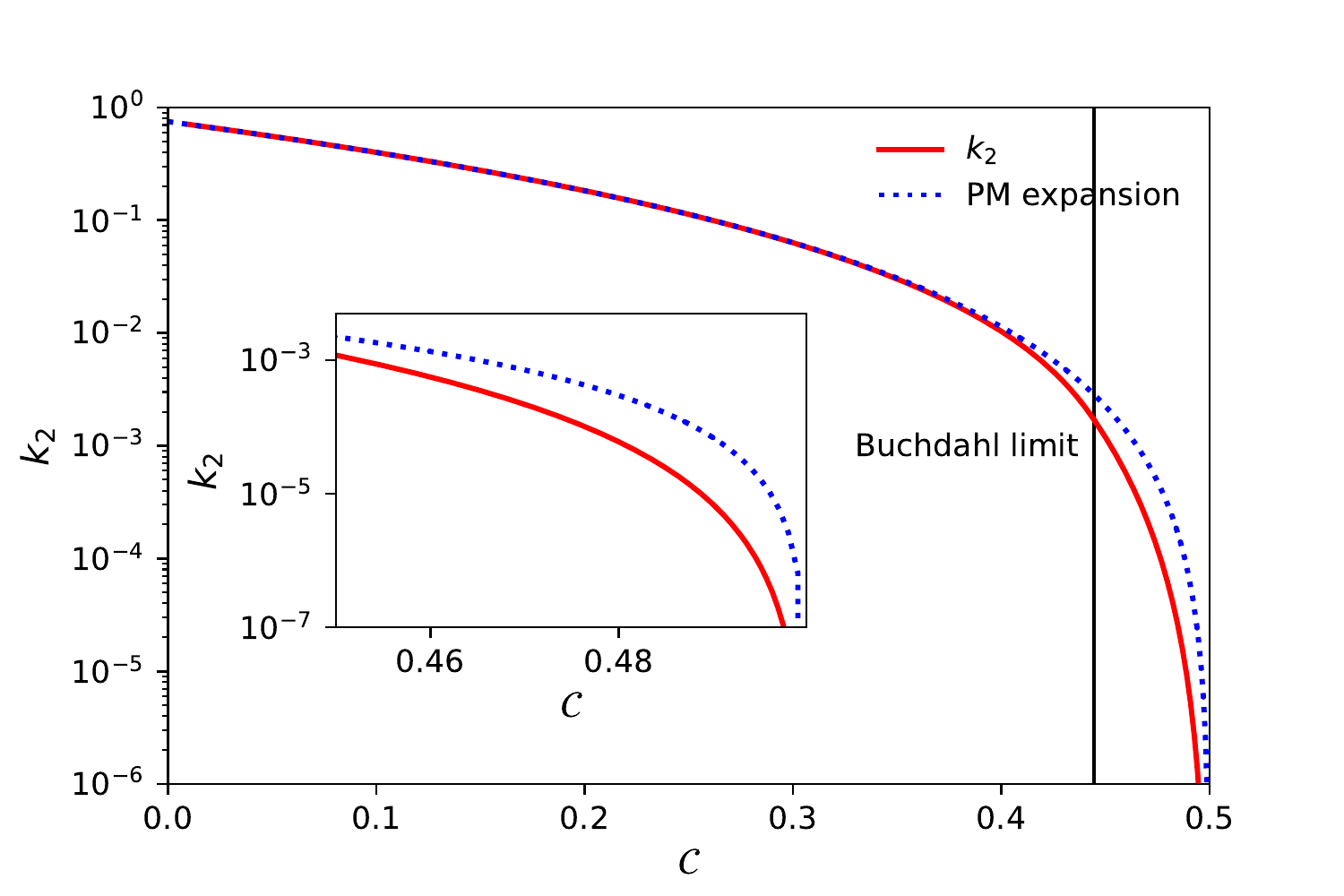}
\end{center}
\caption{Same as Fig. \ref{Fig:1}, but in log-scale to emphasize the approach to the Schwarzschild limit. For comparison, we also show the Post-Minkowskian approximation for constant density stars presented in \citep{Chan:2014tva}. In the Newtonian limit both approaches are in good agreement. In the strong gravity zone, near the Buchdahl limit, the deviations start to grow as expected.} 
\label{Fig:3}
\end{figure}

Finally, in Fig.~\ref{Fig:3} we show in log-scale the tidal Love number, as a function of the compactness, and we also include the Post-Minkowskian (PM) expansion for constant density stars introduced by \cite{Chan:2014tva}. Note that for configurations with low compactness (Newtonian limit), the PM approximation is in good agreement with our results. However, when the compactness approaches the Buchdahl limit, the differences between both results increase. This difference is expected considering that near the Buchdahl limit we are entering into the strong gravity zone, thus the PM expansion is not a good approximation; in the strong gravity regime, we require the full general relativistic computation. In reference \citep{Chirenti:2020bas}, figure 4 shows a similar log-scale for $k_2$, together with the PM approximation. After the publication of that paper, the author found that the plot is marred due to a typo in the PM expansion. However, this error does not affect the results of $k_2$.

\section{Discussion}\label{conclus}%%%%%%%%%%%%%%%%%%%%%%%%%%%%%%%%%%%%%%%

The Schwarzschild interior solution with constant density, or Schwarzschild star, remains as an interesting and simple example of an exact solution to Einstein's equations for a perfect fluid. One of the most remarkable features of this model only discovered recently, is that it allows for the compactness to cross beyond the Buchdahl limit. Moreover, these configurations can approach arbitrarily close the Schwarzschild radius where they become essentially the gravastar.

It has been found that the Love number $k_2$ of the ultracompact Schwarzschild stars is a smooth and continuous function of the compactness, and it approaches zero as the compactness approaches the BH limit 1/2. Thus, one can conclude that the result $k_2\to 0$ is not an exclusive property of BHs, but rather a consequence of the compactness approaching the Schwarzschild limit. 

We found that the Love number, of ultracompact Schwarzschild stars, does not follow the logarithmic behavior conjectured by \citep{Cardoso:2017cfl} for generic ECOs. Instead, we found that $k_2$ decreases rapidly with compactness, following an exponentially decaying behavior. Therefore, we believe that the conclusions drawn by \citep{Cardoso:2017cfl}, regarding the behavior of $k_2$ for ECOS, are too restrictive considering that they studied only 4 types of ECOs, namely, thin-shell gravastar, boson stars, wormholes, and some kind of anisotropic stars. 
%So far, it is unclear how based only on these few examples, these authors concluded that the logarithmic behaviour for the Love number is a generic feature of any ECO.

%Acknowledgements are created using the command \ack:
\ack%%%%%%%%%%%%%%%%%%%%%%%%%%%%%%%%%%%%%%%%%%%%%%%%%%%%%%%%%%%%%%%%%%%%%%%
The author acknowledges the support of the Institute of Physics in Opava and its Research Centre for Theoretical Physics and Astrophysics.
% Contributors involved in `Vyzkumny zamer' can use macro \InstResCode
% instead of specifying the alphanumerical code explicitly:
%The present work was supported by the Czech Grant \InstResCode.

% Here we specify the basename of the bibliography database file,
% in this case \jobname=ragsamp:
%\newpage
\bibliography{references}

\end{document}